\documentclass[twocolumn,showpacs,aps,prl,amsmath,amssymb]{revtex4-1}
\usepackage[dvips]{graphicx}
\usepackage{subfigure}
\usepackage{color}
\usepackage{epsfig}
\usepackage{indentfirst}
\usepackage{multirow}
\usepackage{amsmath}
\usepackage{amssymb}
\usepackage{bm}
\usepackage{fancyhdr}
\usepackage{enumerate}
\usepackage[perpage,symbol]{footmisc}
\usepackage[bookmarksnumbered,pdfstartview=FitH,colorlinks=true,menucolor=black,linkcolor=blue]{hyperref}
\usepackage{xspace}
\definecolor{orange}{RGB}{255,165,0}
\definecolor{chocolate}{RGB}{210,105,30}
\definecolor{purple}{RGB}{255,0,255}

\usepackage{ulem}

\begin{document}
%\linenumbers

%\graphicspath{{figure/}}

%%%%%%%%%%%%%%%%%%%%%%%%%%%%%%%%%%%%%%%%%%%%%%%%%%%%%%%%%%%%%%%%%%%%%%%%%%%
%%                              title
%%%%%%%%%%%%%%%%%%%%%%%%%%%%%%%%%%%%%%%%%%%%%%%%%%%%%%%%%%%%%%%%%%%%%%%%%%%
\title
{
\boldmath \large Measurement of the absolute branching fraction of the inclusive semileptonic $\Lambda_c^+$ decay
}

\author
{
\begin{small}
\begin{center}
M.~Ablikim$^{1}$, M.~N.~Achasov$^{9,d}$, S.~Ahmed$^{14}$, M.~Albrecht$^{4}$, M.~Alekseev$^{53A,53C}$, A.~Amoroso$^{53A,53C}$, F.~F.~An$^{1}$, Q.~An$^{50,40}$, J.~Z.~Bai$^{1}$, Y.~Bai$^{39}$
% ...
\\
\vspace{0.2cm}
(BESIII Collaboration)\\
\vspace{0.2cm} {\it
$^{1}$ Institute of High Energy Physics, Beijing 100049, People's Republic of China\\
% ...
$^{4}$ Bochum Ruhr-University, D-44780 Bochum, Germany\\
% ...
$^{9}$ G.I. Budker Institute of Nuclear Physics SB RAS (BINP), Novosibirsk 630090, Russia\\
% ...
$^{14}$ Helmholtz Institute Mainz, Johann-Joachim-Becher-Weg 45, D-55099 Mainz, Germany\\
% ...
$^{39}$ Southeast University, Nanjing 211100, People's Republic of China\\
$^{40}$ State Key Laboratory of Particle Detection and Electronics, Beijing 100049, Hefei 230026, People's Republic of China\\
% ...
$^{50}$ University of Science and Technology of China, Hefei 230026, People's Republic of China\\
% ...
$^{53}$ (A)University of Turin, I-10125, Turin, Italy; (B)University of Eastern Piedmont, I-15121, Alessandria, Italy; (C)INFN, I-10125, Turin, Italy\\
% ...
$^{d}$ Also at the Novosibirsk State University, Novosibirsk, 630090, Russia\\
% ...
}
\end{center}
\vspace{0.4cm}
\end{small}
}

%\date{\today}

\begin{abstract}
   Using a data sample corresponding to an integrated luminosity of 567 pb$^{-1}$ collected at a center-of-mass energy of $\sqrt{s}=4.6$ GeV with the BESIII detector, we measure the absolute branching fraction of the inclusive semileptonic $\Lambda_c^+$ decay with a double-tag method. We obtain $\mathcal{B}(\Lambda_c^+ \rightarrow X e^+ \nu_e) = (3.95\pm0.34\pm0.09)\%$, where the first uncertainty is statistical and the second systematic. Using the known $\Lambda_c^+$ lifetime and the charge-averaged semileptonic decay width of nonstrange charmed measons ($D^0$ and $D^+$), we obtain the ratio of the inclusive semileptonic decay widths $\Gamma(\Lambda_c^+ \rightarrow X e^+ \nu_e)/\bar{\Gamma}(D\rightarrow X e^+ \nu_e)= 1.26\pm0.12$.
\end{abstract}

\maketitle
%%%%%%%%%%%%%%%%%%%%%%%%%%%%%%%%%%%%%%%%%%%%%%%%%%%%%%%%%%%%%%%%%%%%%%%%%%%
%%                              introduction
%%%%%%%%%%%%%%%%%%%%%%%%%%%%%%%%%%%%%%%%%%%%%%%%%%%%%%%%%%%%%%%%%%%%%%%%%%%
%%                              motivation
Since the first observation of the $\Lambda_c^+$ baryon, the lightest
baryon containing a charm quark, in 1979~\cite{Abrams:1980}, its hadronic decays have been studied extensively. However, information about semileptonic decays of the $\Lambda_c^+$ baryon is sparse~\cite{Gronau:2018, Albrecht:1991, Bergfeld:1994, Ablikim:2015, Ablikim:2017}. The branching fraction of $\Lambda_c^+ \rightarrow \Lambda e^+ \nu_e$ was first measured by the ARGUS collaboration~\cite{Albrecht:1991} and then measured by the CLEO collaboration~\cite{Bergfeld:1994} more than 20 years ago. Recently, the BESIII collaboration measured the absolute branching fraction of $\Lambda_c^+ \rightarrow \Lambda e^+ \nu_e$ to be $(3.63 \pm 0.43)\%$~\cite{Ablikim:2015}. A comparison of this exclusive branching fraction and the inclusive semileptonic decay branching fraction of the $\Lambda_c^+$ baryon will guide searches for new semileptonic decay modes.
The branching fraction of the inclusive semileptonic decay has been
measured previously by the MARK II collaboration 35 years ago, with a
result of $(4.5 \pm 1.7)\%$~\cite{Vella:1982}. The uncertainty is much
larger than that of the exclusive decay. Thus, a more precise
measurement for the inclusive semileptonic decay is required. In
addition, using the known $\Lambda_c^+$ lifetime, the semileptonic
decay width $\Gamma(\Lambda_c^+ \rightarrow X e^+ \nu_e)$, where $X$
refers to any particle system with baryon number one, can be
determined. Comparing $\Gamma(\Lambda_c^+ \rightarrow X e^+ \nu_e)$
with the charge-averaged nonstrange $D$ semileptonic decay width
$\bar{\Gamma}(D \rightarrow X e^+ \nu_e)$, the ratio
$\Gamma(\Lambda_c^+ \rightarrow X e^+ \nu_e)/\bar{\Gamma}(D\rightarrow
X e^+ \nu_e)$ can be obtained. Using current data results in $\Gamma(\Lambda_c^+ \rightarrow X e^+ \nu_e)/\bar{\Gamma}(D\rightarrow X e^+ \nu_e)=1.44\pm0.54$~\cite{PDG:2016, Gronau:2011}.
This ratio is predicted to be 1.67~\cite{Gronau:2011, Rosner:2012} using an effective-quark theory calculation and about 1.2 based on a calculation using the heavy-quark expansion~\cite{Manohar:1994}.
Therefore, a more precise measurement of $\mathcal{B}({\Lambda_c^+} \rightarrow X e^+ \nu_e)$ is desirable to test these theoretical predictions.

In this Letter, we present the first absolute measurement of the branching fraction of the inclusive semileptonic $\Lambda_c^+$ decay using a double-tag method.
This analysis is based on a data sample corresponding to an integrated luminosity of 567 pb$^{-1}$, which is the largest $\Lambda^+_c$ sample taken just above the $\Lambda^+_c\bar \Lambda^-_c$ production threshold collected up to now. The data sample was accumulated at a center-of-mass energy $\sqrt{s}=4.6$ GeV  and recorded with the BESIII detector~\cite{Ablikim:2010-bes3} at the Beijing Electron-Positron Collider II (BEPCII). A detailed description of the BESIII detector can be found in Ref.~\cite{Ablikim:2010-bes3}.

%%%%%%%%%%%%%%%%%%%%%%%%%%%%%%%%%%%%%%%%%%%%%%%%%%%%%%%%%%%%%%%%%%%%%%%%%%%
%%                              Detector
%%%%%%%%%%%%%%%%%%%%%%%%%%%%%%%%%%%%%%%%%%%%%%%%%%%%%%%%%%%%%%%%%%%%%%%%%%%
A GEANT4-based~\cite{Agostinelli:2003} Monte Carlo (MC) simulation is used to estimate the signal efficiency, optimize the selection criteria and understand the backgrounds. In the simulation, the effects of beam-energy spread and initial state radiation (ISR) are incorporated using {\sc kkmc}~\cite{Jadach:2001}, and the final-state radiation (FSR) is modeled by {\sc photos}~\cite{Richter-Was:1993}. The `inclusive' MC samples consist of $\Lambda_c^+ \Bar{\Lambda}_c^-$ pairs, $D_{(s)}^{(*)} \Bar{D}_{(s)}^{(*)}$ pairs, ISR to lower-mass charmonium ($\psi$) states, and continuum QED and QCD processes {$e^+e^- \rightarrow q\bar{q}$ ($q=u, d, s$)}. All known decay modes of {$\Lambda_c^+$}, $D_{(s)}^{(*)}$ and $\psi$ are generated with the branching fractions taken from the Particle Data Group (PDG)~\cite{PDG:2016} by {\sc evtgen}~\cite{EvtGen:2008, EvtGen:2001}, and the remaining unknown decay modes of $\psi$ are generated by {\sc lundcharm}~\cite{CHARM:2000}.  The equivalent luminosities of the simulated data samples are several times that of real data.

%%%%%%%%%%%%%%%%%%%%%%%%%%%%%%%%%%%%%%%%%%%%%%%%%%%%%%%%%%%%%%%%%%%%%%%%%%%
%%                              Analysis method
%%%%%%%%%%%%%%%%%%%%%%%%%%%%%%%%%%%%%%%%%%%%%%%%%%%%%%%%%%%%%%%%%%%%%%%%%%%
Since the data are taken just above the production threshold of $\Lambda_c^+\bar \Lambda^-_c$, no additional hadrons are produced. The double-tag technique, first developed by the MARK III collaboration \cite{Baltrusaitis:1986}, is used to determine the absolute branching fraction of the inclusive semileptonic decay. First, we fully reconstruct one {$\bar{\Lambda}_c^-$} [referred to as the \emph{single-tag} (ST)], and then search for candidates of the signal decay
in the rest of the event that is recoiling against the tagged $\bar \Lambda_c^-$. Hence, the absolute branching fraction of the inclusive semileptonic decay can be measured without knowing the number of $\Lambda_c^+\bar{\Lambda}_c^-$ pairs produced, thus eliminating the related systematic uncertainty from the measurement. The ST candidates are reconstructed  through the decays $\bar{\Lambda}_c^- \rightarrow \bar{p} K^0_S $  and $\bar{\Lambda}_c^- \rightarrow \bar{p} K^+ \pi^-$, which have large branching fractions and low backgrounds. The charge conjugated modes are implied throughout this Letter unless otherwise stated.

%%%%%%%%%%%%%%%%%%%%%%%%%%%%%%%%%%%%%%%%%%%%%%%%%%%%%%%%%%%%%%%%%%%%%%%%%%%
%%                                 ST
%%%%%%%%%%%%%%%%%%%%%%%%%%%%%%%%%%%%%%%%%%%%%%%%%%%%%%%%%%%%%%%%%%%%%%%%%%%
%% Charged track candidates
The charged tracks, except those from $K_S^0$, are required to have a polar angle $\theta$ with respect to the beam direction within the multilayer drift chamber (MDC) acceptance $|\cos\theta|<0.93$, and a distance of closest approach to the interaction point (IP) within 10~cm along the beam direction and 1~cm in the plane transverse to the beam direction.
%% Particle identification
Particle identification (PID) for charged pions, kaons and protons is performed by exploiting time-of-flight (TOF) information and  specific ionization energy loss $dE/dx$ measured by the MDC. The confidence level (C.L.) under each particle hypothesis ($p$, $K$, or $\pi$) is calculated; each charged track is assigned the particle type with the largest PID C.L.
%%  K_s^0
The $K^0_S$ meson candidates are reconstructed from two oppositely charged tracks to which no PID criteria are applied and which are assigned the pion mass hypothesis. The charged tracks from the $K_S^0$ candidate must satisfy $|\cos\theta|<0.93$.  Furthermore, due to the long lifetime of the $K^{0}_{S}$ meson, there is a less stringent criterion on the distance of the closest approach to the IP in the beam direction of less than 20~cm and there is no requirement on the distance of closest approach in the plane transverse to the beam direction.  The invariant mass of the track pair is required to be in the range (0.487, 0.511)~GeV/$c^2$. Furthermore, the $\pi^+\pi^-$ pair is constrained to be consistent with originating from a common decay vertex by means of a vertex fit. In addition, the decay length, which is the distance between the IP and the decay vertex, is required to be larger than twice its resolution.
%% deltaE and mBC

To suppress combinatorial backgrounds, two kinematic variables are used to select the ST candidates. These are the energy difference $\Delta E \equiv E_{\bar{\Lambda}_c^-} -E_{\rm beam}$ and the beam-constrained mass {$M_{\rm BC}=\sqrt{E_{\rm beam}^2/c^4-|\vec{p}_{\bar{\Lambda}_c^-}|^2/c^2}$, where $E_{\rm beam}$ is the beam energy, $E_{\bar{\Lambda}_c^-}$ and $\vec{p}_{\bar{\Lambda}_c^-}$  are the reconstructed energy and three momentum of the ST candidate in the rest frame of the $e^+e^-$ system, respectively. We require $\Delta E$ to be  within $(-3\sigma, 3\sigma)$ of the peak of the $\Delta E$ distribution, where $\sigma$ is the resolution of the $\Delta E$ distribution. Table~\ref{table: summary of ST data} gives the $\Delta E$ requirements for each ST mode. If there are multiple candidates for the same tag mode in a given event, only the combination with the smallest $|\Delta E|$ is retained for further analysis. To determine the ST yields, we apply a fit to the $M_{\rm BC}$ distributions, as shown in Fig.~\ref{fig: mBC data}. In the fits, the signal shape is modeled by the shape obtained from the MC convolved with a Gaussian function that describes the resolution difference between data and MC simulation; the combinatorial background is described by an ARGUS function \cite{Albrecht:1990}. We obtain the ST yields by subtracting the integral of the background function in the signal region {$2.282 <M_{\rm BC}<2.300~{\rm GeV}/c^2$} from the total number of events in the same region. The tails of the $M_{BC}$ distribution above the nominal $\Lambda_c^{+}$ mass are due to the effects of ISR and FSR. The ST yields and the corresponding detection efficiencies are summarized in Table~\ref{table: summary of ST data}.

\begin{table}
  \centering
  \caption{ Summary of $\Delta E$  requirements, detection efficiencies and  ST yields for the different tag modes.}
  \begin{tabular}{lccc}
  \hline
  \hline
  Tag {mode} & $\Delta E$ (MeV) & Efficiency (\%) & Yield\\
  \hline
  $\bar{\Lambda}_c^- \rightarrow \bar{p} K^0_S    $ & $(-21, 19)$ & $56.5\pm0.3$ & $1214\pm36$  \\
  $\bar{\Lambda}_c^- \rightarrow \bar{p} K^+ \pi^-$ & $(-20, 16)$ & $50.1\pm0.1$ & $6092\pm82$  \\
  \hline
  \hline
  \end{tabular}
  \label{table: summary of ST data}
\end{table}

\begin{figure}
  \centering
  \includegraphics[width = 0.45\textwidth]{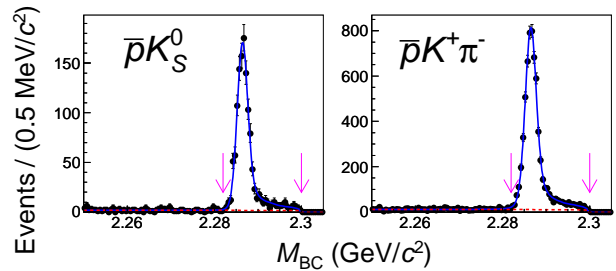}
  \caption{(Color online) $M_{\rm BC}$ distributions for the different ST modes in data. The solid blue line is the total fit, the dashed red line is the background component, and the pink arrows denote the $M_{\rm BC}$ signal region.}
  \label{fig: mBC data}
\end{figure}

%%%%%%%%%%%%%%%%%%%%%%%%%%%%%%%%%%%%%%%%%%%%%%%%%%%%%%%%%%%%%%%%%%%%%%%%%%%
%%                                 DT
%%%%%%%%%%%%%%%%%%%%%%%%%%%%%%%%%%%%%%%%%%%%%%%%%%%%%%%%%%%%%%%%%%%%%%%%%%%
%% method
In the selected ST sample of $\bar{\Lambda}_c^-$ candidates, we  search for charged tracks consistent with being an electron or positron.
To ensure that the charged tracks originate from the IP, the same distance of closest approach selection criteria are used as for the non-$K^{0}_{S}$ daughters of the ST candidates. The track is required to satisfy $|\cos\theta|<0.8$ to ensure that it lies within the acceptance of the barrel of the electromagnetic calorimeter (EMC), which has better energy resolution than the EMC end-caps.
The momentum of the charged track is required to be greater than 200~MeV$/c$, as it is difficult to separate positrons from other hadrons with low momenta. The selected tracks are divided into right-sign (RS) and wrong-sign (WS) samples, where the charge of the RS (WS) track is required to be opposite (equal) to that of the ST candidate.

%% PID
The PID of the selected tracks is implemented with the information of the $dE/dx$, TOF and EMC, and the C.L. under each particle hypothesis ($e$, $\pi$, $K$ or $p$) is calculated. Positron candidates must satisfy $CL(e) > 0.001$ and $CL(e)/(CL(e)+CL(\pi)+CL(K)+CL(p))>0.8$. To further suppress the backgrounds from charged pions, $E_e/p_e > 0.8$ is required, where $E_e$ and $p_e$ are the deposited energy in the EMC and momentum measured by the MDC, respectively. The remaining selected charged tracks are assigned the hadron type corresponding to the highest C.L. that is greater than 0.001. The track is rejected if it does not have a C.L. greater than 0.001 for any hypothesis.

%% A: PID unfolding
The identified positron sample contains  sizable backgrounds from misidentified hadrons.
To evaluate these backgrounds, knowledge of their yields and corresponding misidentification probabilities is required.
The real RS and WS positron yields are determined individually by unfolding the matrix~\cite{A. Hocker:1996, N. E. Adam:2006, D. M. Asner:2010}
\[
\label{equation: PID-unfolding}
\small
\left(
\begin{array}{cccc}
N_{e}^{\rm obs} \\
N_{\pi}^{\rm obs} \\
N_{K}^{\rm obs} \\
N_{p}^{\rm obs}
\end{array}
\right)
=
\left(
\begin{array}{cccc}
P_{e \rightarrow e} & P_{\pi \rightarrow e} & P_{K \rightarrow e} & P_{p \rightarrow e} \\
P_{e \rightarrow \pi} & P_{\pi \rightarrow \pi} & P_{K \rightarrow \pi} & P_{p \rightarrow \pi} \\
P_{e \rightarrow K} & P_{\pi \rightarrow K} & P_{K \rightarrow K} & P_{p \rightarrow K} \\
P_{e \rightarrow p} & P_{\pi \rightarrow p} & P_{K \rightarrow p} & P_{p \rightarrow p}
\end{array}
\right)
\left(
\begin{array}{cccc}
N_{e}^{\rm true} \\
N_{\pi}^{\rm true} \\
N_{K}^{\rm true} \\
N_{p}^{\rm true}
\end{array}
\right),
\]
where $N_{a}^{\rm obs}$ is the observed yield of particle species $a$ ($a$ denotes $e$, $\pi$, $K$ or $p$), $P_{a \rightarrow b}$ is the probability to identify particle $a$ as particle $b$, and $N_{a}^{\rm true}$ is the true yield of particle $a$ in the studied sample.
The elements of the PID efficiency matrix $P_{a \rightarrow b}$ are obtained by studying corresponding control samples selected from data.
The charged pion and proton samples are selected from $J/\psi \rightarrow p \bar{p} \pi^+ \pi^-$ events. The charged kaon and positron samples are selected from $J/\psi \rightarrow K^+  K^- K^+ K^-$ and radiative Bhabha events, respectively.  Due to the different event topologies, the PID efficiency of positrons from $\Lambda_c^+\Bar{\Lambda}_c^-$ pairs (one positron and several hadrons) differs from that from radiative Bhabha events (one electron, one positron and one shower). The relative difference ($\sim$ 4.2\%) is corrected by comparing the positron efficiency obtained from radiative Bhabha MC samples and $\Lambda_c^+\Bar{\Lambda}_c^-$ pair MC samples. No correction to the other elements is implemented. The momentum dependence of the PID efficiency matrix is mostly determined in intervals of 100~MeV/$c$, though some intervals are wider due to limited statistics, as presented in Fig.~\ref{fig: PID efficiency matrix data}.
The muon component is omitted in the unfolding procedure due to its small yields (almost the same as the positron yields), the small mis-PID probability from muon to positron (similar to that from pion to positron, shown in Fig.~\ref{fig: PID efficiency matrix data}) and the negligible effect on the branching fraction measurement. In addition, because the selected pion sample contains the muon component due to their similar PID behaviour in the BESIII detector, the muon component is implicitly taken into account.

%% figure of P_a_a
\begin{figure}
  \centering
  \includegraphics[width = 0.45\textwidth]{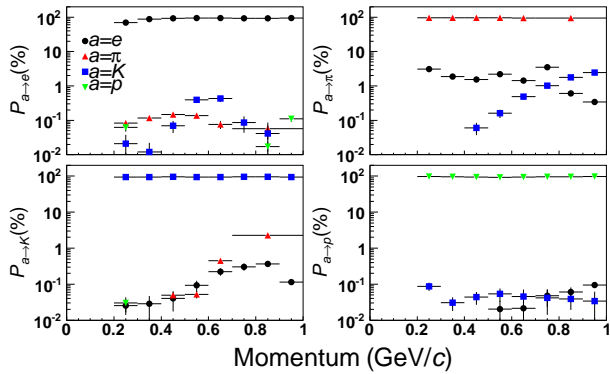}\\
  \caption{(Color online) PID efficiencies obtained from data.}
  \label{fig: PID efficiency matrix data}
\end{figure}

% summarize bkg level
%% B: Tag sideband subtraction
To estimate the contribution from non-$\Lambda^+_c$ decays in the signal region, the unfolded positron yield in the $M_{\rm BC}$ sideband region is scaled by a factor of 0.78 that accounts for the relative amount of background in the sideband and signal regions determined by the fit to the $M_{\rm BC}$ distribution. Since low-background ST modes are used, the contribution from non-$\Lambda_c^+$ decays is small (3.8\%).

%% C: WS subtraction
The RS sample contains primary positrons, which directly originate from $\Lambda_c^+$ decays, and secondary positrons, not directly arising from $\Lambda_c^+$ decays and originating predominantly from $\gamma$ conversions and $\pi^0$ Dalitz decays. Detailed MC studies indicate that the secondary positrons are charge symmetric, hence their yield can be evaluated from the WS positron sample and subtracted from the total RS positron yields. The reliability of the WS subtraction has been validated by MC studies.

%% D: Correction of the tracking efficiency
The tracking efficiency in a given momentum interval, including the track reconstruction efficiency, selection efficiency and resolution effects, is corrected by unfolding the following matrix equation
\begin{equation}\label{function: tracking efficiency}
  {N^{\rm true}_{i} = \sum_{j} T(i|j) N^{\rm pro}_{j},}
\end{equation}
where the tracking efficiency matrix $T(i|j)$ describes the probability of positrons produced in the $j$-th momentum interval to be reconstructed in the $i$-th momentum interval, $N^{\rm pro}_{j}$ is the number of  primary positrons produced in the $j$-th momentum interval and $N^{\rm true}_{i}$ is the true yield of positron reconstructed in the $i$-th momentum interval. The tracking efficiency matrix is obtained by studying the positron MC sample selected from $\Lambda_c^+$ semileptonic events. After this procedure, we obtain the efficiency-corrected positron momentum spectrum above 200~MeV$/c$ in the laboratory frame. Table~\ref{table: summary of signal yields} summarizes the positron yields obtained after each correction step.

\begin{table}
  \centering
  \caption{ Positron yields in data after each procedure. The uncertainties are statistical.}
  \begin{tabular}{lcc}
  \hline
  \hline
%                                        & \multicolumn{2}{c}{$\Lambda_c^+ \rightarrow X e^+ \nu_e$}   \\
%                        &  {RS}               & {WS}  \\
%
  $\Lambda_c^+ \rightarrow X e^+ \nu_e$                     &  {RS}               & {WS}   \\
  \hline
  Observed yields                      \\
  ~~tag signal region                   & $228.0\pm15.1$            & $26.0\pm5.1$\\
  ~~tag sideband region                 & $11.0\pm3.3$              & {$2.0\pm1.4$} \\

  PID unfolding                                     \\
  ~~tag signal region                   & $250.1\pm17.1$            & $28.3\pm6.2$  \\
  ~~tag sideband region                 & $12.1\pm3.8$              & $1.7\pm1.5$   \\

  Sideband subtraction                  & $240.7\pm17.4$            & $27.0\pm6.3$  \\
  WS subtraction                        & $213.7\pm18.5$            &\\
  Correction of tracking efficiency     & $272.1\pm23.5$            &\\
  %Extrapolation                         & $288.3\pm24.9$            &\\

  \hline
  \hline
  \end{tabular}
  \label{table: summary of signal yields}
\end{table}

%% F: Extrapolation
The fraction of positrons below 200 MeV$/c$ is obtained by fitting the efficiency-corrected positron momentum spectrum with the sum of the spectra of the exclusive decay channels (Table~\ref{table: four}), as shown in Fig.~\ref{fig: extrapolation}. In the fit, the branching fraction of each component is allowed to vary within the given uncertainty.
%% explain the fit procedure...
From the fit, we obtain the fraction of positrons below 200 MeV$/c$ to be $(5.6 \pm 1.5)\%$, where the uncertainty is systematic derived from variations of the fit assumptions. The branching fraction of the inclusive semileptonic decay of the $\Lambda_c^+$ baryon is then calculated with
\begin{equation}
\mathcal{B}(\Lambda_c^+ \rightarrow X e^+ \nu_e)=\frac{N^{\rm pro}(p_e>200~{\rm MeV}/c)}{N_{\rm tag}[1-f(p_e<200~{\rm MeV}/c)]},
\end{equation}
where $N^{\rm pro}(p_e>200~{\rm MeV}/c)$ is the yield of positrons with momentum $p_e$ above 200 MeV$/c$ after the correction of the tracking efficiency, $N_{\rm tag}$ is the ST yield and $f(p_e<200~{\rm MeV}/c)$ is the fraction of positron below 200 MeV$/c$. Finally, we obtain  $\mathcal{B}(\Lambda_c^+ \rightarrow X e^+ \nu_e) = (3.95\pm0.34)\%$, where the uncertainty includes only the statistical component of that on the signal and ST yields.

\begin {table}
\caption {$\Lambda_c^+$ semileptonic decays used to extrapolate the positron momentum spectrum. The branching fraction of $\Lambda_c^+\rightarrow \Lambda e^+ \nu_e$ decay is from BESIII measurement~\cite{Ablikim:2015} and the uncertainty of the unobserved decay
channels are 100\%  of the predicted branching fractions. The form factor of $\Lambda_c^+ \rightarrow \Lambda e^+ \nu_e$ decay is from QCD sum rules~\cite{R. S. Marques:1999} and the other two, unobserved, semileptonic decay modes are generated by {\sc pythia}~\cite{PYTHIA: 2001} according to the simple $V-A$ matrix element.}
\label{table: four}
\centering
\begin{tabular} {lcc}
  \hline \hline
  Decay channel                                      &$\mathcal{B}$ (\%)  & Model    \\
  \hline
  $\Lambda_c^+ \rightarrow \Lambda e^+ \nu_e$        & $3.63\pm0.43$~\cite{Ablikim:2015}&  $F_1^V(q^2)=\frac{2.52}{5.09-q^2}$~\cite{R. S. Marques:1999}  \\
  $\Lambda_c^+ \rightarrow \Lambda(1405) e^+ \nu_e$  & $0.38\pm0.38$~\cite{M. Pervin:2005} & {\sc pythia}~\cite{PYTHIA: 2001}        \\
  $\Lambda_c^+ \rightarrow n e^+ \nu_e$              & $0.27\pm0.27$~\cite{M. A. Ivanov:1997}  & {\sc pythia}~\cite{PYTHIA: 2001}     \\
  \hline \hline
\end{tabular}
\end{table}
\begin{figure}
  \centering
  \includegraphics[width=0.45\textwidth]{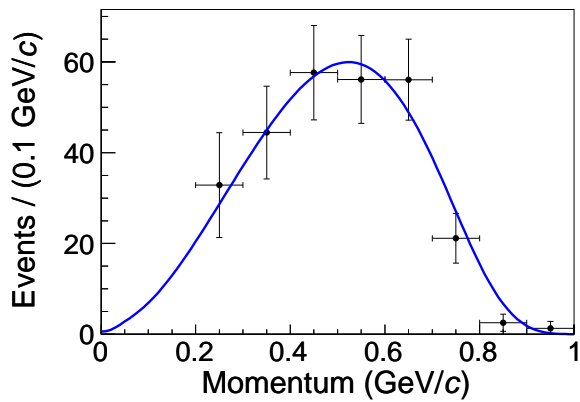}

  \caption{(Color online) Extrapolation of the positron momentum spectrum in the laboratory frame obtained from data, shown as points with error bars. The blue curve shows the extrapolated spectrum.}
  \label{fig: extrapolation}
\end{figure}

%%%%%%%%%%%%%%%%%%%%%%%%%%%%%%%%%%%%%%%%%%%%%%%%%%%%%%%%%%%%%%%%%%%%%%%%%%%
%%                              systematic uncertianty
%%%%%%%%%%%%%%%%%%%%%%%%%%%%%%%%%%%%%%%%%%%%%%%%%%%%%%%%%%%%%%%%%%%%%%%%%%%
The systematic uncertainties in this analysis are listed in Table~\ref{table: summary of systematic uncertainties}.
%% Tag yield
The tag yield systematic uncertainty is estimated to be 1.0\% by using alternative fits to the $M_{\rm BC}$ distribution with different signal shapes, background parameters and fitting ranges.
%% Tracking
The systematic uncertainty related to the tracking efficiency is estimated to be 1.0\% by studying radiative Bhabha events~\cite{Ablikim:2015}.
%% PID
The systematic uncertainty in the positron identification efficiency is estimated by comparing the positron PID efficiencies in different MC simulated semileptonic $\Lambda_c^+$ decays. The largest relative difference of the positron PID efficiency is assigned as the systematic uncertainty.
The uncertainties in the other elements of the PID efficiency matrix are estimated by comparing the matrix elements obtained from $\Lambda_c^+\Bar{\Lambda}_c^-$ pair MC samples with those obtained from radiative Bhabha, $J/\psi \rightarrow p \bar{p} \pi^+ \pi^-$ and $J/\psi \rightarrow K^+K^-K^+K^-$ MC samples.
Adding them in quadrature, we assign 0.9\% as the systematic uncertainty related to PID.
%% Sideband subtraction
The uncertainty associated with the $M_{\rm BC}$ sideband subtraction is estimated to be 0.5\% by using an alternative $M_{\rm BC}$ sideband region.
%% Extrapolation
To estimate the uncertainty in the extrapolation of the positron momentum spectrum, we perform an alternative fit in which the branching fraction of each fit component is unconstrained. In addition, we use an alternative form-factor model and repeat the fit. Adding these effects in quadrature, we attribute 1.5\% as the systematic uncertainty related to the extrapolation procedure.
%% Data and MC statistics
The uncertainty due to limited statistics of data and MC simulation used to determine the PID efficiency matrix and tracking efficiency matrix is estimated by repeating the PID unfolding procedure and correction of tracking efficiency. In each repetition, we vary each element of the PID efficiency matrix and tracking efficiency matrix within the corresponding error. The corresponding systematic uncertainty is derived from $10,000$ independent repetitions and is estimated to be 0.4\%.
%% Sum
Adding all uncertainties in quadrature, the total systematic uncertainty is determined to be 2.3\%.

\begin{table}
  \centering
  \caption{ {Sources of systematic uncertainties.}}
  \begin{tabular}{lc}
    \hline \hline
    Source                 &       Relative uncertainty~(\%)  \\%& Comments\\
    \hline
    Tag yield               &       1.0                       \\%& signal shape, background shape and fit range\\
    Tracking                &       1.0                       \\%& difference between data and MC\\
    PID                     &       0.9                       \\%& efficiency difference of $P_{e\rightarrow e}$ \\
    Sideband subtraction    &       0.5                       \\%& change sideband region\\
    Extrapolation           &       1.5                       \\%& change parameters of $\Lambda_c^+ \rightarrow \Lambda e^+ \nu_e$ and add unobserved channels \\
    Data and MC statistics  &       0.4                       \\%& vary each element in deviation\\
    \hline
    Sum                     &       2.3                       \\
    \hline \hline
  \end{tabular}

  \label {table: summary of systematic uncertainties}
\end{table}

%%%%%%%%%%%%%%%%%%%%%%%%%%%%%%%%%%%%%%%%%%%%%%%%%%%%%%%%%%%%%%%%%%%%%%%%%%%
%%                              result
%%%%%%%%%%%%%%%%%%%%%%%%%%%%%%%%%%%%%%%%%%%%%%%%%%%%%%%%%%%%%%%%%%%%%%%%%%%
The absolute branching fraction of the inclusive semileptonic decays of the $\Lambda_c^+$ baryon is determined to be $\mathcal{B}(\Lambda_c^+ \rightarrow X e^+ \nu_e) = (3.95\pm0.34\pm0.09)\%$, where the first and second uncertainties are statistical and systematic, respectively. Compared with the branching fraction of $\Lambda_c^+ \rightarrow \Lambda e^+ \nu_e$ measured by the BESIII collabration~\cite{Ablikim:2015}, the ratio $\frac{\mathcal{B}(\Lambda_c^+ \rightarrow \Lambda e^+ \nu_e)}{\mathcal{B}(\Lambda_c^+ \rightarrow Xe^+\nu_e)}$ is determined to be $(91.9\pm12.5\pm5.4)\%$, where the systematic uncertainty related to the tracking efficiency of the positron cancels. Using the known $\Lambda_c^+$ lifetime~\cite{PDG:2016}, we obtain the semileptonic decay width $\Gamma(\Lambda_c^+ \rightarrow X e^+ \nu_e)=(1.98\pm0.18)\times 10^{11}~{\rm s}^{-1}$.  Comparing this with the charge-averaged semileptonic decay width of nonstrange charmed measons $\bar{\Gamma}(D\rightarrow X e^+ \nu_e)$~\cite{PDG:2016}, the ratio $\frac{\Gamma(\Lambda_c^+ \rightarrow X e^+ \nu_e)}{\bar{\Gamma}(D\rightarrow X e^+ \nu_e)}$ is determined to be $1.26\pm0.12$. A comparison of the branching fraction and ratio of the semileptonic decay width between experimental measurements and theoretical predictions can be found in Table~\ref{table: where}.
\begin{table}
  \centering
  \caption{ Comparison of the branching fraction (in $10^{-2}$) and ratio of the semileptonic decay width between experimental measurements and theoretical predictions.}
  \label{table: where}
  \begin{tabular}{l c c}
    \hline \hline
    Result &   $\Lambda_c^+ \rightarrow X e^+ \nu_e$   &    $\frac{\Gamma(\Lambda_c^+ \rightarrow X e^+ \nu_e)}{\bar{\Gamma}(D\rightarrow X e^+ \nu_e)}$   \\%& Comments\\
    \hline
    BESIII      &       $3.95\pm0.35$                &   $1.26\pm0.12$      \\
    MARK II~\cite{Vella:1982}                    &       $4.5\pm1.7$                  &   $1.44\pm0.54$      \\
    Effective-quark Method~\cite{Gronau:2011, Rosner:2012}      &                                     & 1.67  \\
    Heavy-quark Expansion~\cite{Manohar:1994}     &                                    &   $1.2$        \\
%    A.~V.~Manohar {\it et al} ~\cite{Manohar:1994}     &                                    &   $1.2$        \\
%    M.~Gronau {\it et al}~\cite{Gronau:2011}      &                                     & 1.67  \\

    \hline \hline
  \end{tabular}
\end{table}

%%%%%%%%%%%%%%%%%%%%%%%%%%%%%%%%%%%%%%%%%%%%%%%%%%%%%%%%%%%%%%%%%%%%%%%%%%%
%%                              summary
%%%%%%%%%%%%%%%%%%%%%%%%%%%%%%%%%%%%%%%%%%%%%%%%%%%%%%%%%%%%%%%%%%%%%%%%%%%
%% BF

In summary, by analysing a data sample corresponding to an integrated luminosity of 567 pb$^{-1}$ taken at a center-of-mass energy $\sqrt{s} = 4.6$ GeV, we report the absolute measurement of the inclusive semileptonic $\Lambda_c^+$ decay branching fraction
$\mathcal{B}(\Lambda_c^+ \rightarrow X e^+ \nu_e) = (3.95\pm0.34\pm0.09)\%$.
The uncertainty is reduced by a factor of four compared to the MARK II result~\cite{Vella:1982}.
%% ratio of branching fraction
Based on the BESIII measurements~\cite{Ablikim:2015}, we obtain the ratio of the branching fraction to be $\frac{\mathcal{B}(\Lambda_c^+ \rightarrow \Lambda e^+ \nu_e)}{\mathcal{B}(\Lambda_c^+ \rightarrow Xe^+\nu_e)}=(91.9\pm12.5\pm5.4)\%$.
We also determine the ratio
$\frac{\Gamma(\Lambda_c^+ \rightarrow X e^+ \nu_e)}{\bar{\Gamma}(D\rightarrow X e^+ \nu_e)}= 1.26\pm 0.12$, which restricts different models as given in Table~\ref{table: where}.}

The BESIII collaboration thanks the staff of BEPCII and the IHEP computing center for their strong support. This work is supported in part by National Key Basic Research Program of China under Contract No. 2015CB856700; National Natural Science Foundation of China (NSFC) under Contracts Nos. 11235011, 11335008, 11425524, 11625523, 11635010; the Chinese Academy of Sciences (CAS) Large-Scale Scientific Facility Program; the CAS Center for Excellence in Particle Physics (CCEPP); Joint Large-Scale Scientific Facility Funds of the NSFC and CAS under Contracts Nos. U1332201, U1532257, U1532258; CAS under Contracts Nos. KJCX2-YW-N29, KJCX2-YW-N45; CAS Key Research Program of Frontier Sciences under Contracts Nos. QYZDJ-SSW-SLH003, QYZDJ-SSW-SLH040; 100 Talents Program of CAS; National 1000 Talents Program of China; INPAC and Shanghai Key Laboratory for Particle Physics and Cosmology; German Research Foundation DFG under Contracts Nos. Collaborative Research Center CRC 1044, FOR 2359; Istituto Nazionale di Fisica Nucleare, Italy; Koninklijke Nederlandse Akademie van Wetenschappen (KNAW) under Contract No. 530-4CDP03; Ministry of Development of Turkey under Contract No. DPT2006K-120470; National Science and Technology fund; The Swedish Research Council; U. S. Department of Energy under Contracts Nos. DE-FG02-05ER41374, DE-SC-0010118, DE-SC-0010504, DE-SC-0012069; University of Groningen (RuG) and the Helmholtzzentrum fuer Schwerionenforschung GmbH (GSI), Darmstadt; WCU Program of National Research Foundation of Korea under Contract No. R32-2008-000-10155-0.

%%%%%%%%%%%%%%%%%%%%%%%%%%%%%%%%%%%%%%%%%%%%%%%%%%%%%%%%%%%%%%%%%%%%%%%%%%%
%%                              bib
%%%%%%%%%%%%%%%%%%%%%%%%%%%%%%%%%%%%%%%%%%%%%%%%%%%%%%%%%%%%%%%%%%%%%%%%%%%

\end{document}